\documentclass[aps, prd,  preprint, superscriptaddress,nofootinbib]{revtex4-1}

\usepackage[utf8]{inputenc}
\usepackage{hyperref}
\usepackage[dvipsnames]{xcolor}
\usepackage{here}
\usepackage{epsfig}
\usepackage{adjustbox}
\usepackage{epstopdf}
\usepackage{soul}
\usepackage{graphicx}
\usepackage{verbatim}
\usepackage{rotating}
\usepackage{graphicx}
\usepackage{amsfonts}
\usepackage{amsmath}
\usepackage{amssymb}
\usepackage{amsthm}
\usepackage[mathscr]{eucal}
\usepackage{bbold}
\usepackage{braket}
\usepackage{color}
\usepackage{comment}
\usepackage{dsfont}
\usepackage{enumitem}
\usepackage{framed}
\usepackage[normalem]{ulem}
\usepackage{tensor}
\usepackage{thmtools}
\usepackage{thm-restate}
\usepackage{ulem}
\usepackage{tikz}

\newcommand{\TeV}{\mathinner{\mathrm{TeV}}}
\newcommand{\GeV}{\mathinner{\mathrm{GeV}}}
\newcommand{\fn}[2]{\mathinner{#1\mathopen{\left(#2\right)}}}
\newcommand{\cref}[1]{Ref.~\cite{#1}}
\newcommand{\fref}[1]{Figure~\ref{#1}}
\newcommand{\tref}[1]{TABLE~\ref{#1}}
\newcommand{\sref}[1]{Section~\ref{#1}}
\newcommand{\eq}[1]{Eq.~(\ref{#1})}

\definecolor{darkYellow}{RGB}{191,191,45}
\definecolor{darkGreen}{RGB}{108,192,45}
\definecolor{darkCyan}{RGB}{108,190,191}

\begin{document}

\title{Stability of the electroweak vacuum with respect to vacuum tunneling to the Komatsu vacuum in the cMSSM}


\author{Hyukjung Kim}
\email{curi951007@gmail.com}
\affiliation{Department of Physics, KAIST, Daejeon 305-701, South Korea} 

\author{Ewan D. Stewart}
\affiliation{Department of Physics, Izmir Institute of Technology, Gulbahce, Urla 35430, Izmir, Turkiye}

\author{Heeseung Zoe}
\email{heeseungzoe@iyte.edu.tr}
\affiliation{Department of Physics, Izmir Institute of Technology, Gulbahce, Urla 35430, Izmir, Turkiye}


\begin{abstract}
We investigate the stability of the electroweak vacuum with respect to vacuum tunneling to the Komatsu vacuum, which exists when $m_L^2 + m_{H_u}^2<0$, in the cMSSM. Employing the numerical tools \texttt{SARAH}, \texttt{SPheno} and \texttt{CosmoTransitions}, we scan and constrain the parameter space of the cMSSM up to 10 TeV. Regions excluded due to having a vacuum tunneling half-life less than the age of the observable universe are concentrated near the regions where the electroweak vacuum is tachyonic and are more stringent at smaller $m_0$, larger and negative $A_0$, and larger $\tan\beta$. New excluded regions, which satisfy $m_h \simeq 125 \text{GeV}$, are found.
\end{abstract}

\maketitle

\section{Introduction}
The Minimal Supersymmetric Standard Model (MSSM) \cite{Martin:1997ns, Aitchison:2005cf, Csaki:1996ks} is one of the most compelling models for physics beyond the Standard Model, as its minimal combination of the Standard Model and supersymmetry not only provides a solution to the hierarchy problem but also predictions that have been confirmed by experiments. First, assuming gauge coupling unification and a supersymmetry mass scale in the range of $10^2$ to $10^4$ GeV, the MSSM correctly \cite{Amaldi:1991cn, Martens:2010nm} predicted \cite{Einhorn:1981sx, Dimopoulos:1981zb} a relation between the Standard Model gauge couplings. Second, the MSSM has consistently predicted the Higgs mass $m_h \lesssim 130$ GeV since well before the LHC \cite{Hempfling:1993qq, Degrassi:2001yf, Brignole:2002bz}. Combined with the experimental bound of $m_h \gtrsim$ 114 GeV from the Large Electron–Positron Collider (LEP) \cite{LEPWorkingGroupforHiggsbosonsearches:2003ing}, the MSSM thus predicted the Higgs boson mass within a $10\%$ range of the value observed at the LHC \cite{{CMS:2012qbp}, ParticleDataGroup:2022pth}.

In the MSSM, the superpartners of the Standard Model particles were expected to be found in the range of $10^2$ to $10^4$ GeV and have been searched for in experiments such as the Large Hadron Collider (LHC). The observation of the Higgs boson mass at $125$ GeV and recent searches have constrained the stop and gluino masses to be above 1 TeV \cite{Draper:2011aa, Heinemeyer:2011aa, Brummer:2012ns, Slavich:2020zjv,ATLAS:2018xpg, CMS:2019agj}. More generally, the LHC has increased the lower bound of the supersymmetric particle masses \cite{ATLAS:2018xpg, CMS:2019agj, Han:2016gvr, Adam:2021rrw, ATLAS:2014hqe, CMS:2016qpc}, and the difference between the electroweak scale and the supersymmetric particle mass scale has become a fine-tuning problem known as the little hierarchy problem.

The Muon g-2 experiment conducted by Brookhaven National Laboratory \cite{Muong-2:2006rrc} and Fermilab \cite{Muong-2:2021ojo, Muong-2:2023cdq} has measured the anomalous magnetic moment of the muon to 20 significant figures. Theoretical estimates of muon g-2 in the Standard Model derived using electron collision data \cite{Aoyama:2020ynm} show a $5 \sigma$ significance difference with the experiment, but lattice QCD simulation of the Standard Model \cite{Borsanyi:2020mff} agrees with the experimental results. Which calculation is correct will determine whether the g-2 experiment is evidence of physics beyond the Standard Model.

Despite a large region of parameter space being ruled out by experiments, the MSSM is still a compelling candidate for physics beyond the Standard Model, especially compared to its alternatives, such as extra dimensions \cite{Arkani-Hamed:1998jmv, Randall:1999ee}, composite Higgs \cite{Miransky:1988xi, Chivukula:1998wd, Hill:2002ap} and cosmological relaxation \cite{Graham:2015cka}, which lack successful predictions. We expect to see evidence for the MSSM in a future collider that can detect sparticles with masses up to 10 TeV \cite{FCC:2018byv, FCC:2018evy, FCC:2018vvp}.

The MSSM has a high dimensional scalar field space, whose potential may have multiple vacua \cite{Drees:1985ie, Kusenko:1996jn, Komatsu:1988mt,  Casas:1995pd, Strumia:1996pr}, which should satisfy the cosmological requirement that the tunneling half-life from the electroweak vacuum to the other vacua should not be much less than the age of the observable universe.

Calculating tunneling in one-dimensional field space is straightforward, but finding the instanton path in multi-dimensional field space is difficult. However, various numerical packages have been developed over the last decade \cite{Wainwright:2011kj, Sato:2019wpo, Guada:2020xnz} and used to constrain MSSM parameters \cite{Camargo-Molina:2013sta, Camargo-Molina:2014pwa, Bechtle:2015nua, Chowdhury:2013dka, Blinov:2013uda, Blinov:2013fta, Chattopadhyay_2014, Hollik_2016, Staub:2018vux, Duan:2018cgb, Hollik:2018wrr, Hollik_2019}. In \cref{Camargo-Molina:2013sta}, the authors scan and constrain the cMSSM parameters up to $4 \TeV$ for vacuum tunneling to vacua formed by $h_u, h_d, \tilde{u}_3, \tilde{\bar{u}}_3, \tilde{e}_3, \tilde{\bar{e}}_3$. In \cref{Camargo-Molina:2014pwa}, the parameter space of the Natural MSSM is constrained using vacuum and thermal tunneling with non-zero $h_u, h_d, \tilde{u}_3, \tilde{\bar{u}}_3$. No points satisfying the Higgs mass constraint $m_h \simeq 125 \GeV$ were excluded by vacuum tunneling, but some were by thermal tunneling. In \cref{Bechtle:2015nua}, the authors assume neutralino dark matter with stau-coannihilation and the theoretical estimate of muon g-2 and find points which best fit with dark matter abundance, muon g-2, and the Higgs mass and decay rate. These points are tested by vacuum tunneling to the Komatsu vacuum, and they are found to be safe. Several other papers \cite{Chowdhury:2013dka, Blinov:2013uda, Blinov:2013fta, Chattopadhyay_2014, Hollik_2016, Staub:2018vux, Duan:2018cgb, Hollik:2018wrr, Hollik_2019} use vacuum tunneling to further reduce the parameter space that survives after making various assumptions and applying constraints. In this paper, we scan and constrain the full cMSSM parameter space up to $10 \TeV$ by vacuum tunneling to the Komatsu vacuum.

The Komatsu vacuum \cite{Komatsu:1988mt,Abel:1998ie} is defined to lie in the direction
\begin{equation}
    \mu H_u \tilde{L}_k = \lambda_d^{ij}\tilde{\bar{d}}_i\tilde{Q}_j\tilde{L}_k + \lambda_e^{ij}\tilde{\bar{e}}_i \tilde{L}_j\tilde{L}_k
\end{equation}
along which the $\mu$ term contribution to $\tilde{L}H_u$'s mass squared is cancelled. At large values of $\tilde{L}H_u$, the D-term constrains $|\tilde{L}| \simeq |H_u|$ and if
\begin{equation}
m_L^2+m_{H_u}^2 < 0  
\end{equation}
the MSSM potential truncated at dimension four becomes unbounded from below. However in reality, the potential simply descends into a deep vacuum, where it is stabilized by higher-order terms such as supersymmetric neutrino mass terms.

In \sref{sec:model}, we present the potential used in this paper and briefly review the calculation of vacuum tunneling. In \sref{sec:methods}, we illustrate the tunneling calculation process, including a brief explanation of the numerical tools: \verb|SARAH| \cite{Staub:2013tta}, \verb|SPheno| \cite{Porod:2011nf} and \verb|CosmoTransitions| \cite{Wainwright:2011kj}. In \sref{sec:results}, we plot the region excluded by vacuum tunneling to the Komatsu vacuum in cMSSM parameter space. We summarize the results and suggest future work in \sref{sec:summary}.

\section{Model}\label{sec:model}

The MSSM superpotential is
\begin{equation}
W = \lambda_u^{ij}{\bar{u}}_i Q_jH_u - \lambda_d^{ij}{\bar{d}}_i {Q}_jH_d -  \lambda_e^{ij}{\bar{e}}_i{L}_jH_d + \mu H_uH_d
\end{equation}
There are two types of Komatsu vacuum: quark Komatsu vacuum $\mu H_u \tilde{L}_k =  \lambda_d^{ij}\tilde{\bar{d}}_i \tilde{Q}_j \tilde{L}_k$ and lepton Komatsu vacuum $\mu H_u \tilde{L}_k =  \lambda_e^{ij} \tilde{\bar{e}}_i \tilde{L}_j \tilde{L}_k$. We have sampled the tunneling rate for the quark Komatsu vacuum and found it to be consistently less than that for the lepton Komatsu vacuum. Thus, we restrict to the lepton Komatsu vacuum
\begin{equation}
 \mu H_u \tilde{L}_k = \lambda_e^{ij}\tilde{\bar{e}}_i \tilde{L}_j \tilde{L}_k  \label{eq:cond2}     
\end{equation}

Note that the $\fn{SU}{2}$ product $\tilde{L}_j\tilde{L}_k=0$ if $j = k$, so we need to consider at least two lepton generations. We neglect the off-diagonal Yukawa couplings and reduce the field space by the following three criteria. First, we build the potential relevant to the electroweak and lepton Komatsu vacua by using  $H_u, H_d, \tilde{L}$ and $\tilde{\bar{e}}$, and setting $\tilde{Q}=\tilde{\bar{u}}=\tilde{\bar{d}}=0$. 
Second, we choose the combination of lepton generations which gives the largest tunneling rate. When the Yukawa coupling $\lambda_e^{ij}$ is large, $\tilde{\bar{e}}_i\tilde{L}_j$ can be small but still cancel the $\mu$ contribution to $\tilde{L}H_u$'s mass squared, resulting in a saddle point close to the origin and a large tunneling rate. Hence, we use $\lambda_{e}^{33}\tilde{\bar{e}}_3\tilde{L}_3$ and set $\tilde{\bar{e}}_1=\tilde{\bar{e}}_2=0$. On the other hand, a larger Yukawa coupling renormalises the corresponding slepton  mass squared to smaller values at low energy, increasing the tunneling rate. Hence we set $\tilde{L}_1= 0$ and use $\tilde{L}_2$.
Third, we use $\fn{SU}{2}$ gauge freedom to set $h_u^+=0$. Then we set $\tilde{e}_2=\tilde{\nu}_3=h_d^-=0$. There are no destabilizing linear terms for these fields, but it is not easy to determine whether the sign of the quadratic terms is always positive, which would be required to fully justify this assumption. Thus, the field configuration we use in this paper is
\begin{eqnarray}
\begin{aligned}
H_u = 
\begin{pmatrix}
0\\
h_u
\end{pmatrix}
\begin{matrix}
,\ 
\end{matrix}
H_d = \begin{pmatrix}
h_d\\
0
\end{pmatrix}
\begin{matrix}
,\ 
\end{matrix}
\tilde{L}_2 = &\begin{pmatrix}
\tilde{\nu}_2 \\
0
\end{pmatrix}
, \\
&
\tilde{L}_3 = \begin{pmatrix}
0 \\
\tilde{e}_3
\end{pmatrix}
\begin{matrix}
,\ 
\end{matrix}
\tilde{\bar{e}}_3=\tilde{\bar{e}}_3 \\
\end{aligned}
\end{eqnarray}
and other fields zero.

At tree-level our potential is 
\begin{equation}
\begin{aligned}
V_{\mathrm{tree}} &= m_{H_u}^2{|h_u|}^2+m_{H_d}^2{|h_d|}^2
+m_{L_2}^2{|\tilde{\nu}_2|}^2+m_{L_3}^2{|\tilde{e}_3|}^2
+m_{\bar{e}_3}^2{|\tilde{\bar{e}}_3|}^2 \\
&+{\lbrack}(a_{e}^{33}\tilde{\bar{e}}_3\tilde{e}_3h_d-B{\mu}h_uh_d)+\mathrm{c.c.}\rbrack \\
&+|\mu h_d|^2+{|\lambda_{e}^{33}\tilde{\bar{e}}_3h_d|}^2+{|{\mu}h_u-\lambda_{e}^{33}\tilde{\bar{e}}_3\tilde{e}_3|}^2 \\
&+\dfrac{1}{8}g_1^2{({|h_u|}^2-{|h_d|}^2-{|\tilde{\nu}_2|}^2-{|\tilde{e}_3|}^2+2{|\tilde{\bar{e}}_3|}^2)}^2 \\
&+ \dfrac{1}{8}g_2^2{({|h_u|}^2-{|h_d|}^2-{|\tilde{\nu}_2|}^2+{|\tilde{e}_3|}^2)}^2
\end{aligned}
\end{equation}

\begin{figure*}[!ht]
\centering
\includegraphics[scale=0.15]{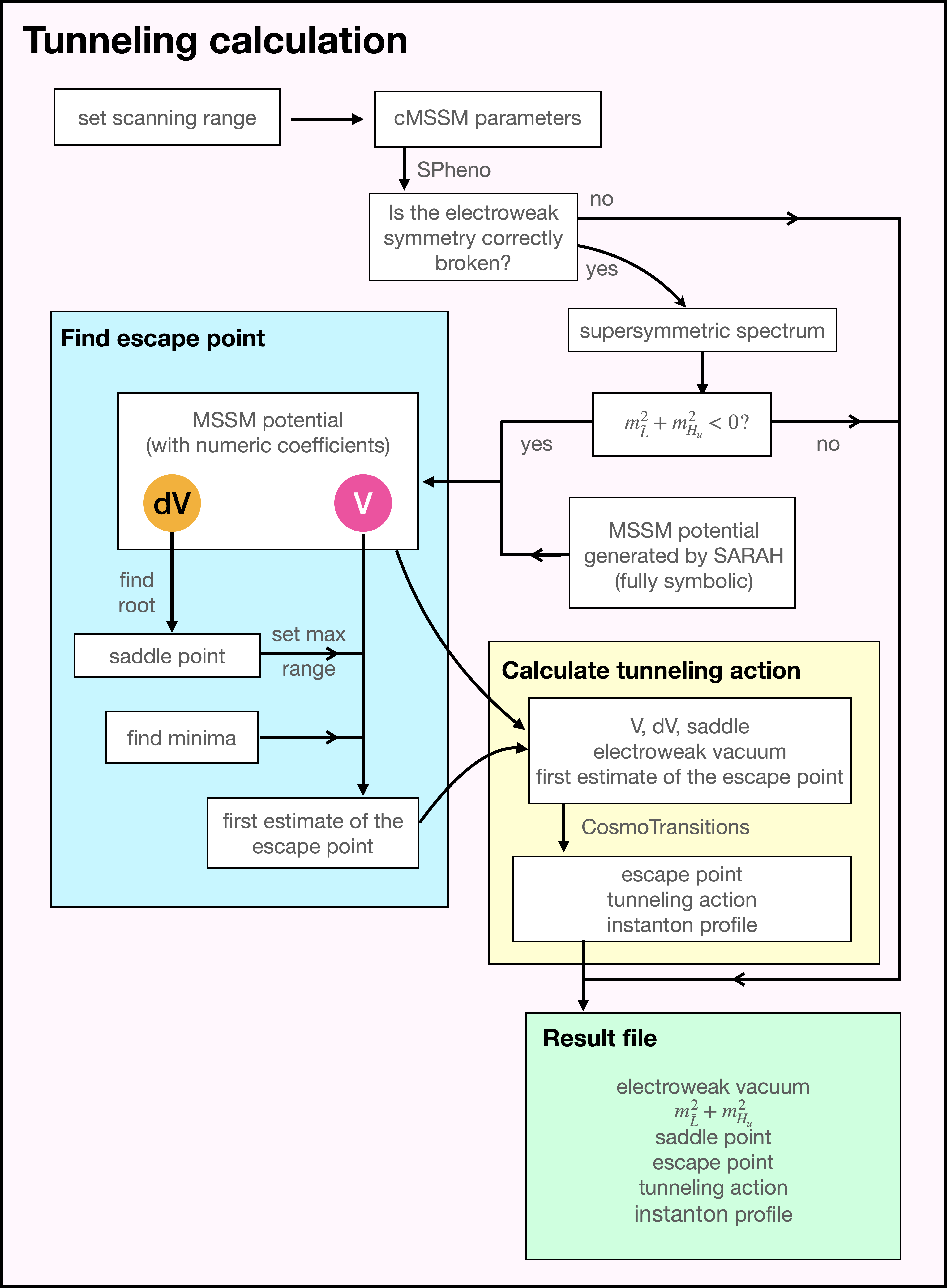}
\caption{Flowchart for the tunneling calculation. $\mathrm{dV} = ( \partial V/\partial h_u, \partial V/\partial h_d, \partial V/\partial \tilde{\nu}_2, \partial V /\partial \tilde{e}_3, \partial V /\partial \tilde{\bar{e}}_3)$ is the gradient of the potential with respect to the field space used in this paper.}
\label{tunneling_process}
\end{figure*}

The tunneling rate to the Komatsu vacuum is calculated by the well-known instanton method \cite{Coleman:1977py}, which gives a tunneling rate
\begin{equation}\label{tunneling_rate}
    \Gamma = Ae^{-S_E}
\end{equation} 
where the prefactor A comes from the measure of the path integral and includes fluctuations around the instanton, and $S_E$ is the Euclidean action given by 
\begin{equation}
    S_E=2\pi^2 \int_{0}^{\infty}\rho^3 d\rho \left[\dfrac{1}{2}h_{ab}\dfrac{d \phi^a}{d\rho}\dfrac{d \phi^b}{d\rho}+ {\fn{V}{\phi}} \right]
\end{equation}
where $\rho$ is the radial distance in Euclidean space. The field space vector
\begin{equation}
\phi = (h_u, h_d, \tilde{\nu}_2, \tilde{e}_3, \tilde{\bar{e}}_3)
\end{equation}
and $h_{ab}$ is the metric on field space. The equation of motion for the instanton tunneling from false vacuum $\phi_{\mathrm{f}}$ to true vacuum $\phi_{\mathrm{t}}$ is
\begin{equation}
    \dfrac{d^2\phi^a}{d\rho^2}+\dfrac{3}{\rho}\dfrac{d\phi^a}{d\rho}=h^{ab}\dfrac{\partial V}{\partial \phi^b}
\end{equation}
with the boundary conditions $ \dfrac{d \phi}{d \rho}(0) =0$, $\phi(0) = \phi_{\mathrm{t}}$ and $\phi(\infty)=\phi_{\mathrm{f}}$.

\section{Method}\label{sec:methods}

\subsection{Numerical tools}

We use \verb|SARAH| \cite{Staub:2013tta} to generate the one-loop corrected potential and source file for \verb|SPheno| \cite{Porod:2011nf}. We calculate the supersymmetric particle spectrum with \verb|SPheno|, including masses and Yukawa couplings at a two-loop level. We use \verb|CosmoTransitions| \cite{Wainwright:2011kj} to calculate the multi-field tunneling. It decomposes the instanton equation into parallel and perpendicular to the path and searches for the instanton path, which is the solution for both parallel and perpendicular equations.

\subsection{Tunneling to the Komatsu vacuum}

The calculation process is illustrated in \fref{tunneling_process}. We work in the context of the cMSSM \cite{Kane:1993td}, which simplifies the MSSM parameters into five parameters: the universal scalar mass $m_0$, the ratio between the MSSM Higgs vacuum expectation values $\tan\beta$, the universal gaugino mass $m_{1/2}$, and the universal trilinear coupling $A_0$, as well as the sign of $\mu$. The scanning range for the cMSSM parameters is shown in \tref{scanning_range}.

\begin{table}[h]
\centering
\begin{tabular}{|c|c|}\hline
Parameter & Range\\ \hline
$m_0/\TeV \; $& $ \{ 0.1, 0.2,0.5,1,2,5,10 \} $\\ \hline
$\tan\beta$& \; $\{10,20,30,40,50\} $\;  \\ \hline
\; $\log_{10}(m_{1/2}/\TeV)$ \;& $[ -1, 1]$ \\ \hline
\; $\log_{10}(|A_0|/\TeV)$\;& $[-1, 1]$\\ \hline
sign of $\mu$ & $+$ \; \\ \hline
\end{tabular}
\caption{Scanning range in cMSSM parameter space}
\label{scanning_range}
\end{table}

We start by setting the scanning range of the cMSSM parameters. \verb|SPheno| checks whether the electroweak symmetry is correctly broken and generates a SUSY spectrum. Next, the existence of the Komatsu vacuum is checked by the sign of $m_{L}^2+m_{H_u}^2$. If the Komatsu vacuum exists, we move on to find the escape point.

We look for the escape point of the bounce solution for potentials unbounded from below in the following manner. First, we find the saddle point\footnote{In addition to the saddle point leading to the Komatsu vacuum, there was another saddle point leading to a vacuum with non-zero $h_u, h_d, \tilde{e}_3, \tilde{\bar{e}}_3$. However, this vacuum has either a potential energy higher than the electroweak vacuum or a tunneling rate lower than to the Komatsu vacuum.} numerically and set  $\phi_* = 5 \times \max \left\{h_u, h_d, \tilde{\nu}_2, \tilde{e}_3, \tilde{\bar{e}}_3\right\}_{\rm saddle}$  to choose the initial scanning range. We find the minimum of the potential within the range $\phi^a < \phi_*$ for each $a$ by using the \verb|minimize| function in \verb|Scipy|. If the electroweak vacuum is found as the minimum, we continuously extend the scanning range to $\phi^a < n \phi_*$ for some positive integer $n$ until we find a minimum within the range $\phi^a < n \phi_*$ in the direction of the Komatsu vacuum with potential energy lower than the electroweak vacuum, which we take as a first estimate of the escape point. Finally, we calculate the bounce action with \verb|CosmoTransitions| and save the results, including the location of the electroweak vacuum, the escape point and the value of the tunneling action.

\begin{figure*}[!htbp]
\begin{adjustbox}{addcode={\begin{minipage}{\width}}{\caption{Constraints from vacuum tunneling to the lepton Komatsu vacuum in cMSSM parameter space. Left to right: $\tan\beta = 10, 20, 30, 40, 50 $; bottom to top: $m_0 = 0.1, 0.2, 0.5, 1, 2, 5, 10$ TeV. \fcolorbox{red}{red}{\rule{0pt}{3pt}\rule{3pt}{0pt}} \ tachyonic electroweak vacuum; \fcolorbox{yellow}{yellow}{\rule{0pt}{3pt}\rule{3pt}{0pt}}, \fcolorbox{darkYellow}{darkYellow}{\rule{0pt}{3pt}\rule{3pt}{0pt}} \ $S<410$; \  \fcolorbox{green}{green}{\rule{0pt}{3pt}\rule{3pt}{0pt}}, \fcolorbox{darkGreen}{darkGreen}{\rule{0pt}{3pt}\rule{3pt}{0pt}}\ $S>410$; \  \fcolorbox{cyan}{cyan}{\rule{0pt}{3pt}\rule{3pt}{0pt}}, \fcolorbox{darkCyan}{darkCyan}{\rule{0pt}{3pt}\rule{3pt}{0pt}}\ no Komatsu vacuum; \fcolorbox{blue}{blue}{\rule{0pt}{3pt}\rule{3pt}{0pt}}\ no electroweak symmetry breaking; \fcolorbox{yellow}{yellow}{\rule{0pt}{3pt}\rule{3pt}{0pt}}, \fcolorbox{green}{green}{\rule{0pt}{3pt}\rule{3pt}{0pt}}, \fcolorbox{cyan}{cyan}{\rule{0pt}{3pt}\rule{3pt}{0pt}} $m_h = 125.25 \pm 1 \GeV$;\ \fcolorbox{darkYellow}{darkYellow}{\rule{0pt}{3pt}\rule{3pt}{0pt}}, \fcolorbox{darkGreen}{darkGreen}{\rule{0pt}{3pt}\rule{3pt}{0pt}}, \fcolorbox{darkCyan}{darkCyan}{\rule{0pt}{3pt}\rule{3pt}{0pt}} $m_h \neq 125.25 \pm 1 \GeV$.}\label{fig:plots_full}\end{minipage}},rotate=90,center}
\includegraphics[scale=0.14]{./tlp_tunneling_action.png}
\end{adjustbox}
\end{figure*}

\begin{figure}[htbp]
\includegraphics[scale=0.155]{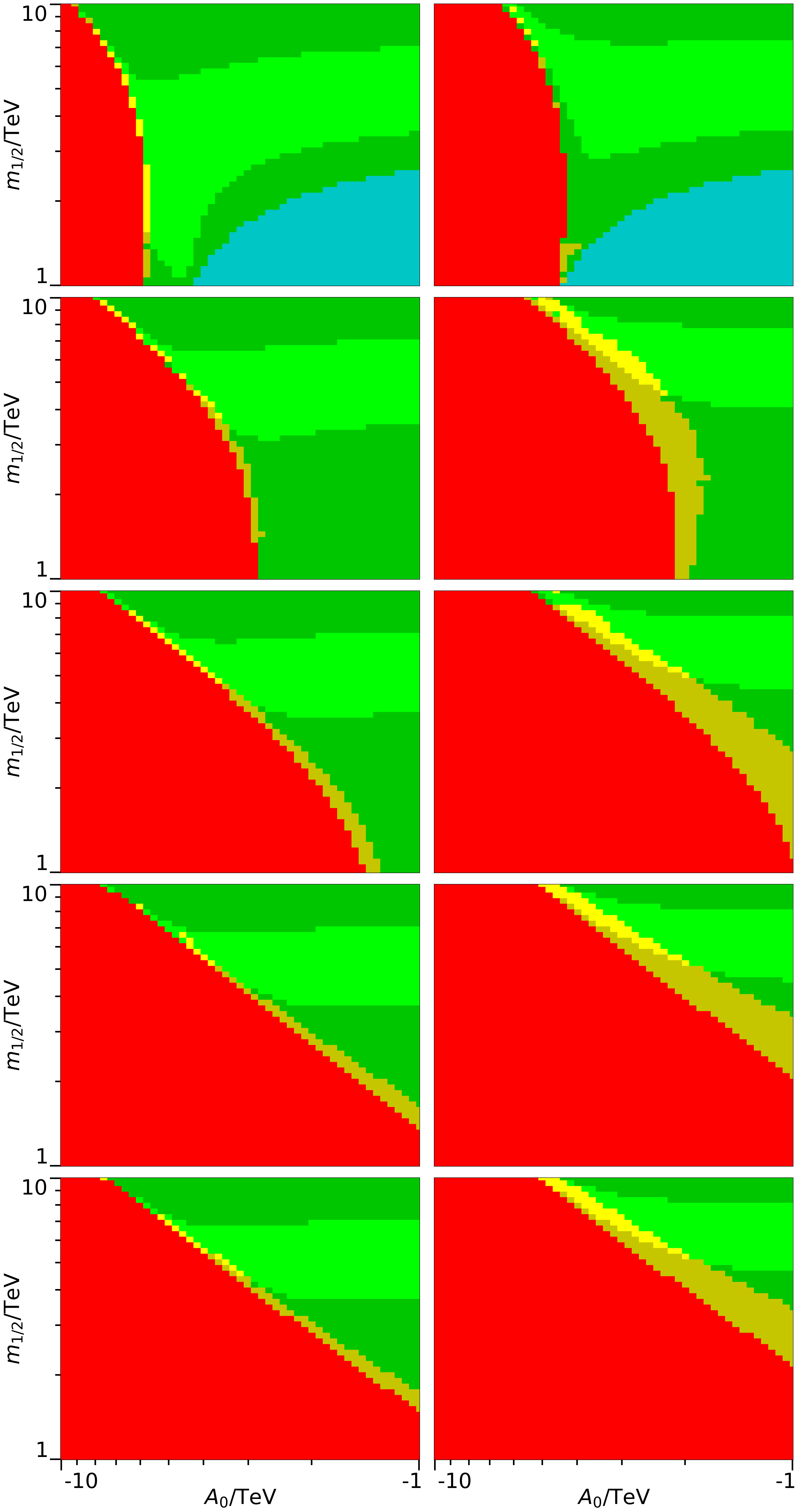}
\caption{Enlargement of the region satisfying the Higgs mass constraint but excluded by tunneling. Left to right: $\tan\beta = 40, 50$; bottom to top: $m_0 = 0.1, 0.2, 0.5, 1, 2$ TeV. Color representation is the same as in \fref{fig:plots_full}. }
\label{fig:plots_const_tlp}
\end{figure}

\section{Results}\label{sec:results}

The constraints on the cMSSM parameter space derived by considering vacuum tunneling from the electroweak vacuum to the lepton Komatsu vacuum are plotted in \fref{fig:plots_full}. We set the prefactor A in \eq{tunneling_rate} to $\TeV^4$, and the action threshold for determining dangerous tunneling to 410, which corresponds to a tunneling half-life of the order of the age of the observable universe. We set the spread of the Higgs mass constraint as $1 \GeV$, which is the theoretical uncertainty of the Higgs mass calculation in $\verb|SPheno|$ \cite{Bahl_2020, Staub_2017}. Note that we do not assume neutralino dark matter, and without this assumption the LHC exclusion bounds do not further constrain the parameter space after the Higgs mass constraint has been applied.

There are four notable trends in the tunneling constraints. First, regions excluded by tunneling (yellow) are located near the regions where the electroweak vacuum is tachyonic (red). The sparticles are close to tachyonic near the red regions, so the height of the saddle is lower, reducing the action. Second, the constraints are stronger for smaller $m_0$ because the mass squareds are smaller, lowering the height of the saddle. Third, the constraints are stronger for larger $\tan\beta$ since $\lambda_e$ is larger allowing $\tilde{\bar{e}}_i\tilde{L}_j$ to be smaller, but still cancel the $\mu$ contribution to mass squared of $\tilde{L}H_u$, resulting in a saddle point located nearer the origin and a smaller action. Fourth, the constraints are stronger for large and negative $A_0$. For negative $A_0$, the trilinear couplings are larger in magnitude at low energy, which renormalises the mass squareds to be smaller at low energy, lowering the height of the saddle.

Our main findings are as follows. For $\tan\beta\leq 30$, only a few points near the tachyonic region are excluded by tunneling and they don't satisfy the Higgs mass constraint. However, for $\tan\beta = 40, 50$, we find a region, enlarged in \fref{fig:plots_const_tlp}, satisfying the Higgs mass constraint but excluded by tunneling, which has not been reported previously.

We have also scanned using the one-loop corrected potential but the tunneling results are not significantly different compared to those of tree-level potential tunneling, see Appendix A. We did not consider the full one-loop corrections to the tunneling \cite{Guada:2020ihz}.

\section{Summary}\label{sec:summary}

In this paper, we have constrained the cMSSM parameter space up to $10 \TeV$ by requiring the vacuum tunneling half-life to the lepton Komatsu vacuum to be greater than the age of the observable universe. The results in \fref{fig:plots_full} show that the tunneling constraints are significant only at $\tan\beta =40,50$ and $m_0 < 2 \TeV$, and the excluded regions are situated near where the electroweak vacuum is tachyonic. \fref{fig:plots_const_tlp} enlarges the region satisfying the Higgs mass constraint but excluded by tunneling. The results may give some indication of what happens in other models since the cMSSM parameter space is fairly large. Constraints from vacuum tunneling are not strong but, unlike potentially stronger constraints from thermal tunneling, they do not dependent on the cosmological history.


\appendix
\section{Tunneling results using the one-loop corrected potential}
In Figures 4 and 5, we show the tunneling constraints obtained using the one-loop corrected potential rather than the tree-level potential. However, the full one-loop corrections to the tunneling are not included so it is not clear how meaningful these results are, apart from suggesting that the full one-loop corrections may also be small. \begin{figure*}[!htbp]
\begin{adjustbox}{addcode={\begin{minipage}{\width}}{\caption{Constraints from vacuum tunneling to the lepton Komatsu vacuum in cMSSM parameter space using the one-loop corrected potential. Left to right: $\tan\beta = 10, 20, 30, 40, 50 $; bottom to top: $m_0 = 0.1, 0.2, 0.5, 1, 2, 5, 10$ TeV. \fcolorbox{red}{red}{\rule{0pt}{3pt}\rule{3pt}{0pt}} \ tachyonic electroweak vacuum; \fcolorbox{yellow}{yellow}{\rule{0pt}{3pt}\rule{3pt}{0pt}}, \fcolorbox{darkYellow}{darkYellow}{\rule{0pt}{3pt}\rule{3pt}{0pt}} \ $S<410$; \  \fcolorbox{green}{green}{\rule{0pt}{3pt}\rule{3pt}{0pt}}, \fcolorbox{darkGreen}{darkGreen}{\rule{0pt}{3pt}\rule{3pt}{0pt}}\ $S>410$; \  \fcolorbox{cyan}{cyan}{\rule{0pt}{3pt}\rule{3pt}{0pt}}, \fcolorbox{darkCyan}{darkCyan}{\rule{0pt}{3pt}\rule{3pt}{0pt}}\ no Komatsu vacuum; \fcolorbox{blue}{blue}{\rule{0pt}{3pt}\rule{3pt}{0pt}}\ no electroweak symmetry breaking; \fcolorbox{yellow}{yellow}{\rule{0pt}{3pt}\rule{3pt}{0pt}}, \fcolorbox{green}{green}{\rule{0pt}{3pt}\rule{3pt}{0pt}}, \fcolorbox{cyan}{cyan}{\rule{0pt}{3pt}\rule{3pt}{0pt}} $m_h = 125.25 \pm 1 \GeV$;\ \fcolorbox{darkYellow}{darkYellow}{\rule{0pt}{3pt}\rule{3pt}{0pt}}, \fcolorbox{darkGreen}{darkGreen}{\rule{0pt}{3pt}\rule{3pt}{0pt}}, \fcolorbox{darkCyan}{darkCyan}{\rule{0pt}{3pt}\rule{3pt}{0pt}} $m_h \neq 125.25 \pm 1 \GeV$.
      }\label{fig:lcp_plots_full}\end{minipage}},rotate=90,center}
\includegraphics[scale=0.14]{./lcp_tunneling_action.png}
\end{adjustbox}
\end{figure*}

\begin{figure}[htbp]
\includegraphics[scale=0.155]{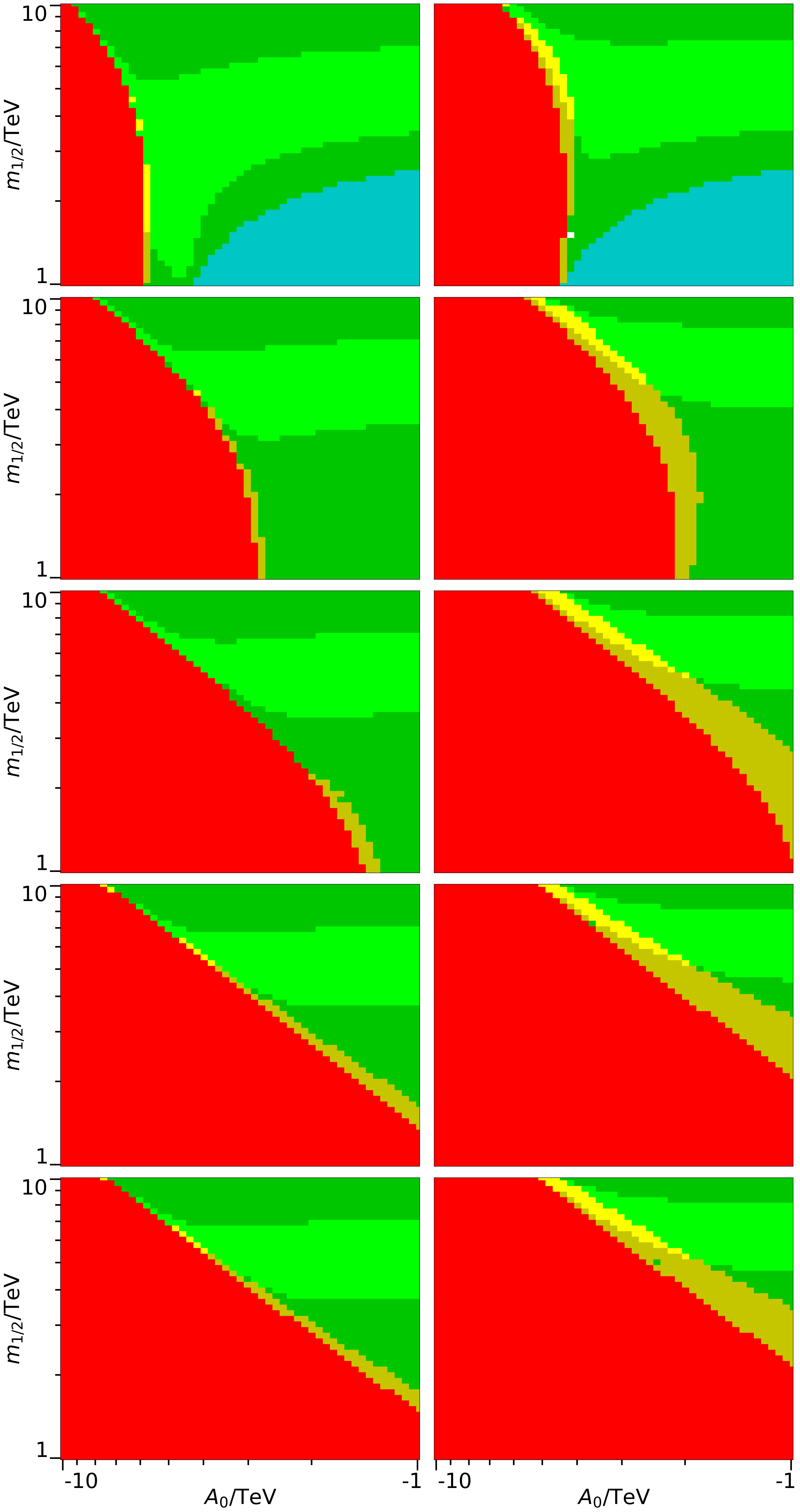}
\caption{Enlargement of the region satisfying the Higgs mass constraint but excluded by tunneling using the one-loop corrected potential. Note that Left to right: $\tan\beta = 40, 50$; bottom to top: $m_0 = 0.1, 0.2, 0.5, 1, 2$ TeV. Color representation is the same as in \fref{fig:lcp_plots_full}. }
\label{fig:lcp_plots_const_tlp}
\end{figure}

\section*{Acknowledgements}
This work is supported by T\"{U}B\.{I}TAK-ARDEB-1001 program under project 123F257. 

\bibliographystyle{apsrev4-1}   
\bibliography{stability}

\begin{thebibliography}{65}%
\makeatletter
\providecommand \@ifxundefined [1]{%
 \@ifx{#1\undefined}
}%
\providecommand \@ifnum [1]{%
 \ifnum #1\expandafter \@firstoftwo
 \else \expandafter \@secondoftwo
 \fi
}%
\providecommand \@ifx [1]{%
 \ifx #1\expandafter \@firstoftwo
 \else \expandafter \@secondoftwo
 \fi
}%
\providecommand \natexlab [1]{#1}%
\providecommand \enquote  [1]{``#1''}%
\providecommand \bibnamefont  [1]{#1}%
\providecommand \bibfnamefont [1]{#1}%
\providecommand \citenamefont [1]{#1}%
\providecommand \href@noop [0]{\@secondoftwo}%
\providecommand \href [0]{\begingroup \@sanitize@url \@href}%
\providecommand \@href[1]{\@@startlink{#1}\@@href}%
\providecommand \@@href[1]{\endgroup#1\@@endlink}%
\providecommand \@sanitize@url [0]{\catcode `\\12\catcode `\$12\catcode `\&12\catcode `\#12\catcode `\^12\catcode `\_12\catcode `\%12\relax}%
\providecommand \@@startlink[1]{}%
\providecommand \@@endlink[0]{}%
\providecommand \url  [0]{\begingroup\@sanitize@url \@url }%
\providecommand \@url [1]{\endgroup\@href {#1}{\urlprefix }}%
\providecommand \urlprefix  [0]{URL }%
\providecommand \Eprint [0]{\href }%
\providecommand \doibase [0]{http://dx.doi.org/}%
\providecommand \selectlanguage [0]{\@gobble}%
\providecommand \bibinfo  [0]{\@secondoftwo}%
\providecommand \bibfield  [0]{\@secondoftwo}%
\providecommand \translation [1]{[#1]}%
\providecommand \BibitemOpen [0]{}%
\providecommand \bibitemStop [0]{}%
\providecommand \bibitemNoStop [0]{.\EOS\space}%
\providecommand \EOS [0]{\spacefactor3000\relax}%
\providecommand \BibitemShut  [1]{\csname bibitem#1\endcsname}%
\let\auto@bib@innerbib\@empty
\bibitem [{\citenamefont {Martin}(1998)}]{Martin:1997ns}%
  \BibitemOpen
  \bibfield  {author} {\bibinfo {author} {\bibfnamefont {S.~P.}\ \bibnamefont {Martin}},\ }\href {\doibase 10.1142/9789812839657_0001} {\bibfield  {journal} {\bibinfo  {journal} {Adv. Ser. Direct. High Energy Phys.}\ }\textbf {\bibinfo {volume} {18}},\ \bibinfo {pages} {1} (\bibinfo {year} {1998})},\ \Eprint {http://arxiv.org/abs/hep-ph/9709356} {arXiv:hep-ph/9709356} \BibitemShut {NoStop}%
\bibitem [{\citenamefont {Aitchison}(2005)}]{Aitchison:2005cf}%
  \BibitemOpen
  \bibfield  {author} {\bibinfo {author} {\bibfnamefont {I.}~\bibnamefont {Aitchison}},\ }\href@noop {} {\emph {\bibinfo {title} {{Supersymmetry and the MSSM: An Elementary introduction}}}}\ (\bibinfo {year} {2005})\ \Eprint {http://arxiv.org/abs/hep-ph/0505105} {arXiv:hep-ph/0505105} \BibitemShut {NoStop}%
\bibitem [{\citenamefont {Csaki}(1996)}]{Csaki:1996ks}%
  \BibitemOpen
  \bibfield  {author} {\bibinfo {author} {\bibfnamefont {C.}~\bibnamefont {Csaki}},\ }\href {\doibase 10.1142/S021773239600062X} {\bibfield  {journal} {\bibinfo  {journal} {Mod. Phys. Lett. A}\ }\textbf {\bibinfo {volume} {11}},\ \bibinfo {pages} {599} (\bibinfo {year} {1996})},\ \Eprint {http://arxiv.org/abs/hep-ph/9606414} {arXiv:hep-ph/9606414} \BibitemShut {NoStop}%
\bibitem [{\citenamefont {Amaldi}\ \emph {et~al.}(1991)\citenamefont {Amaldi}, \citenamefont {de~Boer},\ and\ \citenamefont {Furstenau}}]{Amaldi:1991cn}%
  \BibitemOpen
  \bibfield  {author} {\bibinfo {author} {\bibfnamefont {U.}~\bibnamefont {Amaldi}}, \bibinfo {author} {\bibfnamefont {W.}~\bibnamefont {de~Boer}}, \ and\ \bibinfo {author} {\bibfnamefont {H.}~\bibnamefont {Furstenau}},\ }\href {\doibase 10.1016/0370-2693(91)91641-8} {\bibfield  {journal} {\bibinfo  {journal} {Phys. Lett. B}\ }\textbf {\bibinfo {volume} {260}},\ \bibinfo {pages} {447} (\bibinfo {year} {1991})}\BibitemShut {NoStop}%
\bibitem [{\citenamefont {Martens}\ \emph {et~al.}(2010)\citenamefont {Martens}, \citenamefont {Mihaila}, \citenamefont {Salomon},\ and\ \citenamefont {Steinhauser}}]{Martens:2010nm}%
  \BibitemOpen
  \bibfield  {author} {\bibinfo {author} {\bibfnamefont {W.}~\bibnamefont {Martens}}, \bibinfo {author} {\bibfnamefont {L.}~\bibnamefont {Mihaila}}, \bibinfo {author} {\bibfnamefont {J.}~\bibnamefont {Salomon}}, \ and\ \bibinfo {author} {\bibfnamefont {M.}~\bibnamefont {Steinhauser}},\ }\href {\doibase 10.1103/PhysRevD.82.095013} {\bibfield  {journal} {\bibinfo  {journal} {Phys. Rev. D}\ }\textbf {\bibinfo {volume} {82}},\ \bibinfo {pages} {095013} (\bibinfo {year} {2010})},\ \Eprint {http://arxiv.org/abs/1008.3070} {arXiv:1008.3070 [hep-ph]} \BibitemShut {NoStop}%
\bibitem [{\citenamefont {Einhorn}\ and\ \citenamefont {Jones}(1982)}]{Einhorn:1981sx}%
  \BibitemOpen
  \bibfield  {author} {\bibinfo {author} {\bibfnamefont {M.~B.}\ \bibnamefont {Einhorn}}\ and\ \bibinfo {author} {\bibfnamefont {D.~R.~T.}\ \bibnamefont {Jones}},\ }\href {\doibase 10.1016/0550-3213(82)90502-8} {\bibfield  {journal} {\bibinfo  {journal} {Nucl. Phys. B}\ }\textbf {\bibinfo {volume} {196}},\ \bibinfo {pages} {475} (\bibinfo {year} {1982})}\BibitemShut {NoStop}%
\bibitem [{\citenamefont {Dimopoulos}\ and\ \citenamefont {Georgi}(1981)}]{Dimopoulos:1981zb}%
  \BibitemOpen
  \bibfield  {author} {\bibinfo {author} {\bibfnamefont {S.}~\bibnamefont {Dimopoulos}}\ and\ \bibinfo {author} {\bibfnamefont {H.}~\bibnamefont {Georgi}},\ }\href {\doibase 10.1016/0550-3213(81)90522-8} {\bibfield  {journal} {\bibinfo  {journal} {Nucl. Phys. B}\ }\textbf {\bibinfo {volume} {193}},\ \bibinfo {pages} {150} (\bibinfo {year} {1981})}\BibitemShut {NoStop}%
\bibitem [{\citenamefont {Hempfling}\ and\ \citenamefont {Hoang}(1994)}]{Hempfling:1993qq}%
  \BibitemOpen
  \bibfield  {author} {\bibinfo {author} {\bibfnamefont {R.}~\bibnamefont {Hempfling}}\ and\ \bibinfo {author} {\bibfnamefont {A.~H.}\ \bibnamefont {Hoang}},\ }\href {\doibase 10.1016/0370-2693(94)90948-2} {\bibfield  {journal} {\bibinfo  {journal} {Phys. Lett. B}\ }\textbf {\bibinfo {volume} {331}},\ \bibinfo {pages} {99} (\bibinfo {year} {1994})},\ \Eprint {http://arxiv.org/abs/hep-ph/9401219} {arXiv:hep-ph/9401219} \BibitemShut {NoStop}%
\bibitem [{\citenamefont {Degrassi}\ \emph {et~al.}(2001)\citenamefont {Degrassi}, \citenamefont {Slavich},\ and\ \citenamefont {Zwirner}}]{Degrassi:2001yf}%
  \BibitemOpen
  \bibfield  {author} {\bibinfo {author} {\bibfnamefont {G.}~\bibnamefont {Degrassi}}, \bibinfo {author} {\bibfnamefont {P.}~\bibnamefont {Slavich}}, \ and\ \bibinfo {author} {\bibfnamefont {F.}~\bibnamefont {Zwirner}},\ }\href {\doibase 10.1016/S0550-3213(01)00343-1} {\bibfield  {journal} {\bibinfo  {journal} {Nucl. Phys. B}\ }\textbf {\bibinfo {volume} {611}},\ \bibinfo {pages} {403} (\bibinfo {year} {2001})},\ \Eprint {http://arxiv.org/abs/hep-ph/0105096} {arXiv:hep-ph/0105096} \BibitemShut {NoStop}%
\bibitem [{\citenamefont {Brignole}\ \emph {et~al.}(2002)\citenamefont {Brignole}, \citenamefont {Degrassi}, \citenamefont {Slavich},\ and\ \citenamefont {Zwirner}}]{Brignole:2002bz}%
  \BibitemOpen
  \bibfield  {author} {\bibinfo {author} {\bibfnamefont {A.}~\bibnamefont {Brignole}}, \bibinfo {author} {\bibfnamefont {G.}~\bibnamefont {Degrassi}}, \bibinfo {author} {\bibfnamefont {P.}~\bibnamefont {Slavich}}, \ and\ \bibinfo {author} {\bibfnamefont {F.}~\bibnamefont {Zwirner}},\ }\href {\doibase 10.1016/S0550-3213(02)00748-4} {\bibfield  {journal} {\bibinfo  {journal} {Nucl. Phys. B}\ }\textbf {\bibinfo {volume} {643}},\ \bibinfo {pages} {79} (\bibinfo {year} {2002})},\ \Eprint {http://arxiv.org/abs/hep-ph/0206101} {arXiv:hep-ph/0206101} \BibitemShut {NoStop}%
\bibitem [{\citenamefont {Barate}\ \emph {et~al.}(2003)\citenamefont {Barate} \emph {et~al.}}]{LEPWorkingGroupforHiggsbosonsearches:2003ing}%
  \BibitemOpen
  \bibfield  {author} {\bibinfo {author} {\bibfnamefont {R.}~\bibnamefont {Barate}} \emph {et~al.} (\bibinfo {collaboration} {LEP Working Group for Higgs boson searches, ALEPH, DELPHI, L3, OPAL}),\ }\href {\doibase 10.1016/S0370-2693(03)00614-2} {\bibfield  {journal} {\bibinfo  {journal} {Phys. Lett. B}\ }\textbf {\bibinfo {volume} {565}},\ \bibinfo {pages} {61} (\bibinfo {year} {2003})},\ \Eprint {http://arxiv.org/abs/hep-ex/0306033} {arXiv:hep-ex/0306033} \BibitemShut {NoStop}%
\bibitem [{\citenamefont {Chatrchyan}\ \emph {et~al.}(2012)\citenamefont {Chatrchyan} \emph {et~al.}}]{CMS:2012qbp}%
  \BibitemOpen
  \bibfield  {author} {\bibinfo {author} {\bibfnamefont {S.}~\bibnamefont {Chatrchyan}} \emph {et~al.} (\bibinfo {collaboration} {CMS}),\ }\href {\doibase 10.1016/j.physletb.2012.08.021} {\bibfield  {journal} {\bibinfo  {journal} {Phys. Lett. B}\ }\textbf {\bibinfo {volume} {716}},\ \bibinfo {pages} {30} (\bibinfo {year} {2012})},\ \Eprint {http://arxiv.org/abs/1207.7235} {arXiv:1207.7235 [hep-ex]} \BibitemShut {NoStop}%
\bibitem [{\citenamefont {Workman}\ \emph {et~al.}(2022)\citenamefont {Workman} \emph {et~al.}}]{ParticleDataGroup:2022pth}%
  \BibitemOpen
  \bibfield  {author} {\bibinfo {author} {\bibfnamefont {R.~L.}\ \bibnamefont {Workman}} \emph {et~al.} (\bibinfo {collaboration} {Particle Data Group}),\ }\href {\doibase 10.1093/ptep/ptac097} {\bibfield  {journal} {\bibinfo  {journal} {PTEP}\ }\textbf {\bibinfo {volume} {2022}},\ \bibinfo {pages} {083C01} (\bibinfo {year} {2022})}\BibitemShut {NoStop}%
\bibitem [{\citenamefont {Draper}\ \emph {et~al.}(2012)\citenamefont {Draper}, \citenamefont {Meade}, \citenamefont {Reece},\ and\ \citenamefont {Shih}}]{Draper:2011aa}%
  \BibitemOpen
  \bibfield  {author} {\bibinfo {author} {\bibfnamefont {P.}~\bibnamefont {Draper}}, \bibinfo {author} {\bibfnamefont {P.}~\bibnamefont {Meade}}, \bibinfo {author} {\bibfnamefont {M.}~\bibnamefont {Reece}}, \ and\ \bibinfo {author} {\bibfnamefont {D.}~\bibnamefont {Shih}},\ }\href {\doibase 10.1103/PhysRevD.85.095007} {\bibfield  {journal} {\bibinfo  {journal} {Phys. Rev. D}\ }\textbf {\bibinfo {volume} {85}},\ \bibinfo {pages} {095007} (\bibinfo {year} {2012})},\ \Eprint {http://arxiv.org/abs/1112.3068} {arXiv:1112.3068 [hep-ph]} \BibitemShut {NoStop}%
\bibitem [{\citenamefont {Heinemeyer}\ \emph {et~al.}(2012)\citenamefont {Heinemeyer}, \citenamefont {Stal},\ and\ \citenamefont {Weiglein}}]{Heinemeyer:2011aa}%
  \BibitemOpen
  \bibfield  {author} {\bibinfo {author} {\bibfnamefont {S.}~\bibnamefont {Heinemeyer}}, \bibinfo {author} {\bibfnamefont {O.}~\bibnamefont {Stal}}, \ and\ \bibinfo {author} {\bibfnamefont {G.}~\bibnamefont {Weiglein}},\ }\href {\doibase 10.1016/j.physletb.2012.02.084} {\bibfield  {journal} {\bibinfo  {journal} {Phys. Lett. B}\ }\textbf {\bibinfo {volume} {710}},\ \bibinfo {pages} {201} (\bibinfo {year} {2012})},\ \Eprint {http://arxiv.org/abs/1112.3026} {arXiv:1112.3026 [hep-ph]} \BibitemShut {NoStop}%
\bibitem [{\citenamefont {Brummer}\ \emph {et~al.}(2012)\citenamefont {Brummer}, \citenamefont {Kraml},\ and\ \citenamefont {Kulkarni}}]{Brummer:2012ns}%
  \BibitemOpen
  \bibfield  {author} {\bibinfo {author} {\bibfnamefont {F.}~\bibnamefont {Brummer}}, \bibinfo {author} {\bibfnamefont {S.}~\bibnamefont {Kraml}}, \ and\ \bibinfo {author} {\bibfnamefont {S.}~\bibnamefont {Kulkarni}},\ }\href {\doibase 10.1007/JHEP08(2012)089} {\bibfield  {journal} {\bibinfo  {journal} {JHEP}\ }\textbf {\bibinfo {volume} {08}},\ \bibinfo {pages} {089} (\bibinfo {year} {2012})},\ \Eprint {http://arxiv.org/abs/1204.5977} {arXiv:1204.5977 [hep-ph]} \BibitemShut {NoStop}%
\bibitem [{\citenamefont {Slavich}\ \emph {et~al.}(2021)\citenamefont {Slavich} \emph {et~al.}}]{Slavich:2020zjv}%
  \BibitemOpen
  \bibfield  {author} {\bibinfo {author} {\bibfnamefont {P.}~\bibnamefont {Slavich}} \emph {et~al.},\ }\href {\doibase 10.1140/epjc/s10052-021-09198-2} {\bibfield  {journal} {\bibinfo  {journal} {Eur. Phys. J. C}\ }\textbf {\bibinfo {volume} {81}},\ \bibinfo {pages} {450} (\bibinfo {year} {2021})},\ \Eprint {http://arxiv.org/abs/2012.15629} {arXiv:2012.15629 [hep-ph]} \BibitemShut {NoStop}%
\bibitem [{\citenamefont {Aaboud}\ \emph {et~al.}(2018)\citenamefont {Aaboud} \emph {et~al.}}]{ATLAS:2018xpg}%
  \BibitemOpen
  \bibfield  {author} {\bibinfo {author} {\bibfnamefont {M.}~\bibnamefont {Aaboud}} \emph {et~al.} (\bibinfo {collaboration} {ATLAS}),\ }\href {\doibase 10.1103/PhysRevD.98.032008} {\bibfield  {journal} {\bibinfo  {journal} {Phys. Rev. D}\ }\textbf {\bibinfo {volume} {98}},\ \bibinfo {pages} {032008} (\bibinfo {year} {2018})},\ \Eprint {http://arxiv.org/abs/1803.10178} {arXiv:1803.10178 [hep-ex]} \BibitemShut {NoStop}%
\bibitem [{\citenamefont {Sirunyan}\ \emph {et~al.}(2019)\citenamefont {Sirunyan} \emph {et~al.}}]{CMS:2019agj}%
  \BibitemOpen
  \bibfield  {author} {\bibinfo {author} {\bibfnamefont {A.~M.}\ \bibnamefont {Sirunyan}} \emph {et~al.} (\bibinfo {collaboration} {CMS}),\ }\href {\doibase 10.1140/epjc/s10052-019-6926-x} {\bibfield  {journal} {\bibinfo  {journal} {Eur. Phys. J. C}\ }\textbf {\bibinfo {volume} {79}},\ \bibinfo {pages} {444} (\bibinfo {year} {2019})},\ \Eprint {http://arxiv.org/abs/1901.06726} {arXiv:1901.06726 [hep-ex]} \BibitemShut {NoStop}%
\bibitem [{\citenamefont {Han}\ \emph {et~al.}(2017)\citenamefont {Han}, \citenamefont {Hikasa}, \citenamefont {Wu}, \citenamefont {Yang},\ and\ \citenamefont {Zhang}}]{Han:2016gvr}%
  \BibitemOpen
  \bibfield  {author} {\bibinfo {author} {\bibfnamefont {C.}~\bibnamefont {Han}}, \bibinfo {author} {\bibfnamefont {K.-i.}\ \bibnamefont {Hikasa}}, \bibinfo {author} {\bibfnamefont {L.}~\bibnamefont {Wu}}, \bibinfo {author} {\bibfnamefont {J.~M.}\ \bibnamefont {Yang}}, \ and\ \bibinfo {author} {\bibfnamefont {Y.}~\bibnamefont {Zhang}},\ }\href {\doibase 10.1016/j.physletb.2017.04.026} {\bibfield  {journal} {\bibinfo  {journal} {Phys. Lett. B}\ }\textbf {\bibinfo {volume} {769}},\ \bibinfo {pages} {470} (\bibinfo {year} {2017})},\ \Eprint {http://arxiv.org/abs/1612.02296} {arXiv:1612.02296 [hep-ph]} \BibitemShut {NoStop}%
\bibitem [{\citenamefont {Adam}\ and\ \citenamefont {Vivarelli}(2022)}]{Adam:2021rrw}%
  \BibitemOpen
  \bibfield  {author} {\bibinfo {author} {\bibfnamefont {W.}~\bibnamefont {Adam}}\ and\ \bibinfo {author} {\bibfnamefont {I.}~\bibnamefont {Vivarelli}},\ }\href {\doibase 10.1142/S0217751X21300222} {\bibfield  {journal} {\bibinfo  {journal} {Int. J. Mod. Phys. A}\ }\textbf {\bibinfo {volume} {37}},\ \bibinfo {pages} {2130022} (\bibinfo {year} {2022})},\ \Eprint {http://arxiv.org/abs/2111.10180} {arXiv:2111.10180 [hep-ex]} \BibitemShut {NoStop}%
\bibitem [{\citenamefont {Aad}\ \emph {et~al.}(2014)\citenamefont {Aad} \emph {et~al.}}]{ATLAS:2014hqe}%
  \BibitemOpen
  \bibfield  {author} {\bibinfo {author} {\bibfnamefont {G.}~\bibnamefont {Aad}} \emph {et~al.} (\bibinfo {collaboration} {ATLAS}),\ }\href {\doibase 10.1103/PhysRevD.90.052008} {\bibfield  {journal} {\bibinfo  {journal} {Phys. Rev. D}\ }\textbf {\bibinfo {volume} {90}},\ \bibinfo {pages} {052008} (\bibinfo {year} {2014})},\ \Eprint {http://arxiv.org/abs/1407.0608} {arXiv:1407.0608 [hep-ex]} \BibitemShut {NoStop}%
\bibitem [{\citenamefont {Khachatryan}\ \emph {et~al.}(2016)\citenamefont {Khachatryan} \emph {et~al.}}]{CMS:2016qpc}%
  \BibitemOpen
  \bibfield  {author} {\bibinfo {author} {\bibfnamefont {V.}~\bibnamefont {Khachatryan}} \emph {et~al.} (\bibinfo {collaboration} {CMS}),\ }\href {\doibase 10.1140/epjc/s10052-016-4292-5} {\bibfield  {journal} {\bibinfo  {journal} {Eur. Phys. J. C}\ }\textbf {\bibinfo {volume} {76}},\ \bibinfo {pages} {460} (\bibinfo {year} {2016})},\ \Eprint {http://arxiv.org/abs/1603.00765} {arXiv:1603.00765 [hep-ex]} \BibitemShut {NoStop}%
\bibitem [{\citenamefont {Bennett}\ \emph {et~al.}(2006)\citenamefont {Bennett} \emph {et~al.}}]{Muong-2:2006rrc}%
  \BibitemOpen
  \bibfield  {author} {\bibinfo {author} {\bibfnamefont {G.~W.}\ \bibnamefont {Bennett}} \emph {et~al.} (\bibinfo {collaboration} {Muon g-2}),\ }\href {\doibase 10.1103/PhysRevD.73.072003} {\bibfield  {journal} {\bibinfo  {journal} {Phys. Rev. D}\ }\textbf {\bibinfo {volume} {73}},\ \bibinfo {pages} {072003} (\bibinfo {year} {2006})},\ \Eprint {http://arxiv.org/abs/hep-ex/0602035} {arXiv:hep-ex/0602035} \BibitemShut {NoStop}%
\bibitem [{\citenamefont {Abi}\ \emph {et~al.}(2021)\citenamefont {Abi} \emph {et~al.}}]{Muong-2:2021ojo}%
  \BibitemOpen
  \bibfield  {author} {\bibinfo {author} {\bibfnamefont {B.}~\bibnamefont {Abi}} \emph {et~al.} (\bibinfo {collaboration} {Muon g-2}),\ }\href {\doibase 10.1103/PhysRevLett.126.141801} {\bibfield  {journal} {\bibinfo  {journal} {Phys. Rev. Lett.}\ }\textbf {\bibinfo {volume} {126}},\ \bibinfo {pages} {141801} (\bibinfo {year} {2021})},\ \Eprint {http://arxiv.org/abs/2104.03281} {arXiv:2104.03281 [hep-ex]} \BibitemShut {NoStop}%
\bibitem [{\citenamefont {Aguillard}\ \emph {et~al.}(2023)\citenamefont {Aguillard} \emph {et~al.}}]{Muong-2:2023cdq}%
  \BibitemOpen
  \bibfield  {author} {\bibinfo {author} {\bibfnamefont {D.~P.}\ \bibnamefont {Aguillard}} \emph {et~al.} (\bibinfo {collaboration} {Muon g-2}),\ }\href@noop {} {\  (\bibinfo {year} {2023})},\ \Eprint {http://arxiv.org/abs/2308.06230} {arXiv:2308.06230 [hep-ex]} \BibitemShut {NoStop}%
\bibitem [{\citenamefont {Aoyama}\ \emph {et~al.}(2020)\citenamefont {Aoyama} \emph {et~al.}}]{Aoyama:2020ynm}%
  \BibitemOpen
  \bibfield  {author} {\bibinfo {author} {\bibfnamefont {T.}~\bibnamefont {Aoyama}} \emph {et~al.},\ }\href {\doibase 10.1016/j.physrep.2020.07.006} {\bibfield  {journal} {\bibinfo  {journal} {Phys. Rept.}\ }\textbf {\bibinfo {volume} {887}},\ \bibinfo {pages} {1} (\bibinfo {year} {2020})},\ \Eprint {http://arxiv.org/abs/2006.04822} {arXiv:2006.04822 [hep-ph]} \BibitemShut {NoStop}%
\bibitem [{\citenamefont {Borsanyi}\ \emph {et~al.}(2021)\citenamefont {Borsanyi} \emph {et~al.}}]{Borsanyi:2020mff}%
  \BibitemOpen
  \bibfield  {author} {\bibinfo {author} {\bibfnamefont {S.}~\bibnamefont {Borsanyi}} \emph {et~al.},\ }\href {\doibase 10.1038/s41586-021-03418-1} {\bibfield  {journal} {\bibinfo  {journal} {Nature}\ }\textbf {\bibinfo {volume} {593}},\ \bibinfo {pages} {51} (\bibinfo {year} {2021})},\ \Eprint {http://arxiv.org/abs/2002.12347} {arXiv:2002.12347 [hep-lat]} \BibitemShut {NoStop}%
\bibitem [{\citenamefont {Arkani-Hamed}\ \emph {et~al.}(1998)\citenamefont {Arkani-Hamed}, \citenamefont {Dimopoulos},\ and\ \citenamefont {Dvali}}]{Arkani-Hamed:1998jmv}%
  \BibitemOpen
  \bibfield  {author} {\bibinfo {author} {\bibfnamefont {N.}~\bibnamefont {Arkani-Hamed}}, \bibinfo {author} {\bibfnamefont {S.}~\bibnamefont {Dimopoulos}}, \ and\ \bibinfo {author} {\bibfnamefont {G.~R.}\ \bibnamefont {Dvali}},\ }\href {\doibase 10.1016/S0370-2693(98)00466-3} {\bibfield  {journal} {\bibinfo  {journal} {Phys. Lett. B}\ }\textbf {\bibinfo {volume} {429}},\ \bibinfo {pages} {263} (\bibinfo {year} {1998})},\ \Eprint {http://arxiv.org/abs/hep-ph/9803315} {arXiv:hep-ph/9803315} \BibitemShut {NoStop}%
\bibitem [{\citenamefont {Randall}\ and\ \citenamefont {Sundrum}(1999)}]{Randall:1999ee}%
  \BibitemOpen
  \bibfield  {author} {\bibinfo {author} {\bibfnamefont {L.}~\bibnamefont {Randall}}\ and\ \bibinfo {author} {\bibfnamefont {R.}~\bibnamefont {Sundrum}},\ }\href {\doibase 10.1103/PhysRevLett.83.3370} {\bibfield  {journal} {\bibinfo  {journal} {Phys. Rev. Lett.}\ }\textbf {\bibinfo {volume} {83}},\ \bibinfo {pages} {3370} (\bibinfo {year} {1999})},\ \Eprint {http://arxiv.org/abs/hep-ph/9905221} {arXiv:hep-ph/9905221} \BibitemShut {NoStop}%
\bibitem [{\citenamefont {Miransky}\ \emph {et~al.}(1989)\citenamefont {Miransky}, \citenamefont {Tanabashi},\ and\ \citenamefont {Yamawaki}}]{Miransky:1988xi}%
  \BibitemOpen
  \bibfield  {author} {\bibinfo {author} {\bibfnamefont {V.~A.}\ \bibnamefont {Miransky}}, \bibinfo {author} {\bibfnamefont {M.}~\bibnamefont {Tanabashi}}, \ and\ \bibinfo {author} {\bibfnamefont {K.}~\bibnamefont {Yamawaki}},\ }\href {\doibase 10.1016/0370-2693(89)91494-9} {\bibfield  {journal} {\bibinfo  {journal} {Phys. Lett. B}\ }\textbf {\bibinfo {volume} {221}},\ \bibinfo {pages} {177} (\bibinfo {year} {1989})}\BibitemShut {NoStop}%
\bibitem [{\citenamefont {Chivukula}\ \emph {et~al.}(1999)\citenamefont {Chivukula}, \citenamefont {Dobrescu}, \citenamefont {Georgi},\ and\ \citenamefont {Hill}}]{Chivukula:1998wd}%
  \BibitemOpen
  \bibfield  {author} {\bibinfo {author} {\bibfnamefont {R.~S.}\ \bibnamefont {Chivukula}}, \bibinfo {author} {\bibfnamefont {B.~A.}\ \bibnamefont {Dobrescu}}, \bibinfo {author} {\bibfnamefont {H.}~\bibnamefont {Georgi}}, \ and\ \bibinfo {author} {\bibfnamefont {C.~T.}\ \bibnamefont {Hill}},\ }\href {\doibase 10.1103/PhysRevD.59.075003} {\bibfield  {journal} {\bibinfo  {journal} {Phys. Rev. D}\ }\textbf {\bibinfo {volume} {59}},\ \bibinfo {pages} {075003} (\bibinfo {year} {1999})},\ \Eprint {http://arxiv.org/abs/hep-ph/9809470} {arXiv:hep-ph/9809470} \BibitemShut {NoStop}%
\bibitem [{\citenamefont {Hill}\ and\ \citenamefont {Simmons}(2003)}]{Hill:2002ap}%
  \BibitemOpen
  \bibfield  {author} {\bibinfo {author} {\bibfnamefont {C.~T.}\ \bibnamefont {Hill}}\ and\ \bibinfo {author} {\bibfnamefont {E.~H.}\ \bibnamefont {Simmons}},\ }\href {\doibase 10.1016/S0370-1573(03)00140-6} {\bibfield  {journal} {\bibinfo  {journal} {Phys. Rept.}\ }\textbf {\bibinfo {volume} {381}},\ \bibinfo {pages} {235} (\bibinfo {year} {2003})},\ \bibinfo {note} {[Erratum: Phys.Rept. 390, 553--554 (2004)]},\ \Eprint {http://arxiv.org/abs/hep-ph/0203079} {arXiv:hep-ph/0203079} \BibitemShut {NoStop}%
\bibitem [{\citenamefont {Graham}\ \emph {et~al.}(2015)\citenamefont {Graham}, \citenamefont {Kaplan},\ and\ \citenamefont {Rajendran}}]{Graham:2015cka}%
  \BibitemOpen
  \bibfield  {author} {\bibinfo {author} {\bibfnamefont {P.~W.}\ \bibnamefont {Graham}}, \bibinfo {author} {\bibfnamefont {D.~E.}\ \bibnamefont {Kaplan}}, \ and\ \bibinfo {author} {\bibfnamefont {S.}~\bibnamefont {Rajendran}},\ }\href {\doibase 10.1103/PhysRevLett.115.221801} {\bibfield  {journal} {\bibinfo  {journal} {Phys. Rev. Lett.}\ }\textbf {\bibinfo {volume} {115}},\ \bibinfo {pages} {221801} (\bibinfo {year} {2015})},\ \Eprint {http://arxiv.org/abs/1504.07551} {arXiv:1504.07551 [hep-ph]} \BibitemShut {NoStop}%
\bibitem [{\citenamefont {Abada}\ \emph {et~al.}(2019{\natexlab{a}})\citenamefont {Abada} \emph {et~al.}}]{FCC:2018byv}%
  \BibitemOpen
  \bibfield  {author} {\bibinfo {author} {\bibfnamefont {A.}~\bibnamefont {Abada}} \emph {et~al.} (\bibinfo {collaboration} {FCC}),\ }\href {\doibase 10.1140/epjc/s10052-019-6904-3} {\bibfield  {journal} {\bibinfo  {journal} {Eur. Phys. J. C}\ }\textbf {\bibinfo {volume} {79}},\ \bibinfo {pages} {474} (\bibinfo {year} {2019}{\natexlab{a}})}\BibitemShut {NoStop}%
\bibitem [{\citenamefont {Abada}\ \emph {et~al.}(2019{\natexlab{b}})\citenamefont {Abada} \emph {et~al.}}]{FCC:2018evy}%
  \BibitemOpen
  \bibfield  {author} {\bibinfo {author} {\bibfnamefont {A.}~\bibnamefont {Abada}} \emph {et~al.} (\bibinfo {collaboration} {FCC}),\ }\href {\doibase 10.1140/epjst/e2019-900045-4} {\bibfield  {journal} {\bibinfo  {journal} {Eur. Phys. J. ST}\ }\textbf {\bibinfo {volume} {228}},\ \bibinfo {pages} {261} (\bibinfo {year} {2019}{\natexlab{b}})}\BibitemShut {NoStop}%
\bibitem [{\citenamefont {Abada}\ \emph {et~al.}(2019{\natexlab{c}})\citenamefont {Abada} \emph {et~al.}}]{FCC:2018vvp}%
  \BibitemOpen
  \bibfield  {author} {\bibinfo {author} {\bibfnamefont {A.}~\bibnamefont {Abada}} \emph {et~al.} (\bibinfo {collaboration} {FCC}),\ }\href {\doibase 10.1140/epjst/e2019-900087-0} {\bibfield  {journal} {\bibinfo  {journal} {Eur. Phys. J. ST}\ }\textbf {\bibinfo {volume} {228}},\ \bibinfo {pages} {755} (\bibinfo {year} {2019}{\natexlab{c}})}\BibitemShut {NoStop}%
\bibitem [{\citenamefont {Drees}\ \emph {et~al.}(1985)\citenamefont {Drees}, \citenamefont {Gluck},\ and\ \citenamefont {Grassie}}]{Drees:1985ie}%
  \BibitemOpen
  \bibfield  {author} {\bibinfo {author} {\bibfnamefont {M.}~\bibnamefont {Drees}}, \bibinfo {author} {\bibfnamefont {M.}~\bibnamefont {Gluck}}, \ and\ \bibinfo {author} {\bibfnamefont {K.}~\bibnamefont {Grassie}},\ }\href {\doibase 10.1016/0370-2693(85)91538-2} {\bibfield  {journal} {\bibinfo  {journal} {Phys. Lett. B}\ }\textbf {\bibinfo {volume} {157}},\ \bibinfo {pages} {164} (\bibinfo {year} {1985})}\BibitemShut {NoStop}%
\bibitem [{\citenamefont {Kusenko}\ \emph {et~al.}(1996)\citenamefont {Kusenko}, \citenamefont {Langacker},\ and\ \citenamefont {Segre}}]{Kusenko:1996jn}%
  \BibitemOpen
  \bibfield  {author} {\bibinfo {author} {\bibfnamefont {A.}~\bibnamefont {Kusenko}}, \bibinfo {author} {\bibfnamefont {P.}~\bibnamefont {Langacker}}, \ and\ \bibinfo {author} {\bibfnamefont {G.}~\bibnamefont {Segre}},\ }\href {\doibase 10.1103/PhysRevD.54.5824} {\bibfield  {journal} {\bibinfo  {journal} {Phys. Rev. D}\ }\textbf {\bibinfo {volume} {54}},\ \bibinfo {pages} {5824} (\bibinfo {year} {1996})},\ \Eprint {http://arxiv.org/abs/hep-ph/9602414} {arXiv:hep-ph/9602414} \BibitemShut {NoStop}%
\bibitem [{\citenamefont {Komatsu}(1988)}]{Komatsu:1988mt}%
  \BibitemOpen
  \bibfield  {author} {\bibinfo {author} {\bibfnamefont {H.}~\bibnamefont {Komatsu}},\ }\href {\doibase 10.1016/0370-2693(88)91441-4} {\bibfield  {journal} {\bibinfo  {journal} {Phys. Lett. B}\ }\textbf {\bibinfo {volume} {215}},\ \bibinfo {pages} {323} (\bibinfo {year} {1988})}\BibitemShut {NoStop}%
\bibitem [{\citenamefont {Casas}\ \emph {et~al.}(1996)\citenamefont {Casas}, \citenamefont {Lleyda},\ and\ \citenamefont {Munoz}}]{Casas:1995pd}%
  \BibitemOpen
  \bibfield  {author} {\bibinfo {author} {\bibfnamefont {J.~A.}\ \bibnamefont {Casas}}, \bibinfo {author} {\bibfnamefont {A.}~\bibnamefont {Lleyda}}, \ and\ \bibinfo {author} {\bibfnamefont {C.}~\bibnamefont {Munoz}},\ }\href {\doibase 10.1016/0550-3213(96)00194-0} {\bibfield  {journal} {\bibinfo  {journal} {Nucl. Phys. B}\ }\textbf {\bibinfo {volume} {471}},\ \bibinfo {pages} {3} (\bibinfo {year} {1996})},\ \Eprint {http://arxiv.org/abs/hep-ph/9507294} {arXiv:hep-ph/9507294} \BibitemShut {NoStop}%
\bibitem [{\citenamefont {Strumia}(1996)}]{Strumia:1996pr}%
  \BibitemOpen
  \bibfield  {author} {\bibinfo {author} {\bibfnamefont {A.}~\bibnamefont {Strumia}},\ }\href {\doibase 10.1016/S0550-3213(96)00554-8} {\bibfield  {journal} {\bibinfo  {journal} {Nucl. Phys. B}\ }\textbf {\bibinfo {volume} {482}},\ \bibinfo {pages} {24} (\bibinfo {year} {1996})},\ \Eprint {http://arxiv.org/abs/hep-ph/9604417} {arXiv:hep-ph/9604417} \BibitemShut {NoStop}%
\bibitem [{\citenamefont {Wainwright}(2012)}]{Wainwright:2011kj}%
  \BibitemOpen
  \bibfield  {author} {\bibinfo {author} {\bibfnamefont {C.~L.}\ \bibnamefont {Wainwright}},\ }\href {\doibase 10.1016/j.cpc.2012.04.004} {\bibfield  {journal} {\bibinfo  {journal} {Comput. Phys. Commun.}\ }\textbf {\bibinfo {volume} {183}},\ \bibinfo {pages} {2006} (\bibinfo {year} {2012})},\ \Eprint {http://arxiv.org/abs/1109.4189} {arXiv:1109.4189 [hep-ph]} \BibitemShut {NoStop}%
\bibitem [{\citenamefont {Sato}(2021)}]{Sato:2019wpo}%
  \BibitemOpen
  \bibfield  {author} {\bibinfo {author} {\bibfnamefont {R.}~\bibnamefont {Sato}},\ }\href {\doibase 10.1016/j.cpc.2020.107566} {\bibfield  {journal} {\bibinfo  {journal} {Comput. Phys. Commun.}\ }\textbf {\bibinfo {volume} {258}},\ \bibinfo {pages} {107566} (\bibinfo {year} {2021})},\ \Eprint {http://arxiv.org/abs/1908.10868} {arXiv:1908.10868 [hep-ph]} \BibitemShut {NoStop}%
\bibitem [{\citenamefont {Guada}\ \emph {et~al.}(2020)\citenamefont {Guada}, \citenamefont {Nemev\v{s}ek},\ and\ \citenamefont {Pintar}}]{Guada:2020xnz}%
  \BibitemOpen
  \bibfield  {author} {\bibinfo {author} {\bibfnamefont {V.}~\bibnamefont {Guada}}, \bibinfo {author} {\bibfnamefont {M.}~\bibnamefont {Nemev\v{s}ek}}, \ and\ \bibinfo {author} {\bibfnamefont {M.}~\bibnamefont {Pintar}},\ }\href {\doibase 10.1016/j.cpc.2020.107480} {\bibfield  {journal} {\bibinfo  {journal} {Comput. Phys. Commun.}\ }\textbf {\bibinfo {volume} {256}},\ \bibinfo {pages} {107480} (\bibinfo {year} {2020})},\ \Eprint {http://arxiv.org/abs/2002.00881} {arXiv:2002.00881 [hep-ph]} \BibitemShut {NoStop}%
\bibitem [{\citenamefont {Camargo-Molina}\ \emph {et~al.}(2013)\citenamefont {Camargo-Molina}, \citenamefont {O'Leary}, \citenamefont {Porod},\ and\ \citenamefont {Staub}}]{Camargo-Molina:2013sta}%
  \BibitemOpen
  \bibfield  {author} {\bibinfo {author} {\bibfnamefont {J.~E.}\ \bibnamefont {Camargo-Molina}}, \bibinfo {author} {\bibfnamefont {B.}~\bibnamefont {O'Leary}}, \bibinfo {author} {\bibfnamefont {W.}~\bibnamefont {Porod}}, \ and\ \bibinfo {author} {\bibfnamefont {F.}~\bibnamefont {Staub}},\ }\href {\doibase 10.1007/JHEP12(2013)103} {\bibfield  {journal} {\bibinfo  {journal} {JHEP}\ }\textbf {\bibinfo {volume} {12}},\ \bibinfo {pages} {103} (\bibinfo {year} {2013})},\ \Eprint {http://arxiv.org/abs/1309.7212} {arXiv:1309.7212 [hep-ph]} \BibitemShut {NoStop}%
\bibitem [{\citenamefont {Camargo-Molina}\ \emph {et~al.}(2014)\citenamefont {Camargo-Molina}, \citenamefont {Garbrecht}, \citenamefont {O'Leary}, \citenamefont {Porod},\ and\ \citenamefont {Staub}}]{Camargo-Molina:2014pwa}%
  \BibitemOpen
  \bibfield  {author} {\bibinfo {author} {\bibfnamefont {J.~E.}\ \bibnamefont {Camargo-Molina}}, \bibinfo {author} {\bibfnamefont {B.}~\bibnamefont {Garbrecht}}, \bibinfo {author} {\bibfnamefont {B.}~\bibnamefont {O'Leary}}, \bibinfo {author} {\bibfnamefont {W.}~\bibnamefont {Porod}}, \ and\ \bibinfo {author} {\bibfnamefont {F.}~\bibnamefont {Staub}},\ }\href {\doibase 10.1016/j.physletb.2014.08.036} {\bibfield  {journal} {\bibinfo  {journal} {Phys. Lett. B}\ }\textbf {\bibinfo {volume} {737}},\ \bibinfo {pages} {156} (\bibinfo {year} {2014})},\ \Eprint {http://arxiv.org/abs/1405.7376} {arXiv:1405.7376 [hep-ph]} \BibitemShut {NoStop}%
\bibitem [{\citenamefont {Bechtle}\ \emph {et~al.}(2016)\citenamefont {Bechtle} \emph {et~al.}}]{Bechtle:2015nua}%
  \BibitemOpen
  \bibfield  {author} {\bibinfo {author} {\bibfnamefont {P.}~\bibnamefont {Bechtle}} \emph {et~al.},\ }\href {\doibase 10.1140/epjc/s10052-015-3864-0} {\bibfield  {journal} {\bibinfo  {journal} {Eur. Phys. J. C}\ }\textbf {\bibinfo {volume} {76}},\ \bibinfo {pages} {96} (\bibinfo {year} {2016})},\ \Eprint {http://arxiv.org/abs/1508.05951} {arXiv:1508.05951 [hep-ph]} \BibitemShut {NoStop}%
\bibitem [{\citenamefont {Chowdhury}\ \emph {et~al.}(2014)\citenamefont {Chowdhury}, \citenamefont {Godbole}, \citenamefont {Mohan},\ and\ \citenamefont {Vempati}}]{Chowdhury:2013dka}%
  \BibitemOpen
  \bibfield  {author} {\bibinfo {author} {\bibfnamefont {D.}~\bibnamefont {Chowdhury}}, \bibinfo {author} {\bibfnamefont {R.~M.}\ \bibnamefont {Godbole}}, \bibinfo {author} {\bibfnamefont {K.~A.}\ \bibnamefont {Mohan}}, \ and\ \bibinfo {author} {\bibfnamefont {S.~K.}\ \bibnamefont {Vempati}},\ }\href {\doibase 10.1007/JHEP02(2014)110} {\bibfield  {journal} {\bibinfo  {journal} {JHEP}\ }\textbf {\bibinfo {volume} {02}},\ \bibinfo {pages} {110} (\bibinfo {year} {2014})},\ \bibinfo {note} {[Erratum: JHEP 03, 149 (2018)]},\ \Eprint {http://arxiv.org/abs/1310.1932} {arXiv:1310.1932 [hep-ph]} \BibitemShut {NoStop}%
\bibitem [{\citenamefont {Blinov}\ and\ \citenamefont {Morrissey}(2013)}]{Blinov:2013uda}%
  \BibitemOpen
  \bibfield  {author} {\bibinfo {author} {\bibfnamefont {N.}~\bibnamefont {Blinov}}\ and\ \bibinfo {author} {\bibfnamefont {D.~E.}\ \bibnamefont {Morrissey}},\ }in\ \href@noop {} {\emph {\bibinfo {booktitle} {{Meeting of the APS Division of Particles and Fields}}}}\ (\bibinfo {year} {2013})\ \Eprint {http://arxiv.org/abs/1309.7397} {arXiv:1309.7397 [hep-ph]} \BibitemShut {NoStop}%
\bibitem [{\citenamefont {Blinov}\ and\ \citenamefont {Morrissey}(2014)}]{Blinov:2013fta}%
  \BibitemOpen
  \bibfield  {author} {\bibinfo {author} {\bibfnamefont {N.}~\bibnamefont {Blinov}}\ and\ \bibinfo {author} {\bibfnamefont {D.~E.}\ \bibnamefont {Morrissey}},\ }\href {\doibase 10.1007/JHEP03(2014)106} {\bibfield  {journal} {\bibinfo  {journal} {JHEP}\ }\textbf {\bibinfo {volume} {03}},\ \bibinfo {pages} {106} (\bibinfo {year} {2014})},\ \Eprint {http://arxiv.org/abs/1310.4174} {arXiv:1310.4174 [hep-ph]} \BibitemShut {NoStop}%
\bibitem [{\citenamefont {Chattopadhyay}\ and\ \citenamefont {Dey}(2014)}]{Chattopadhyay_2014}%
  \BibitemOpen
  \bibfield  {author} {\bibinfo {author} {\bibfnamefont {U.}~\bibnamefont {Chattopadhyay}}\ and\ \bibinfo {author} {\bibfnamefont {A.}~\bibnamefont {Dey}},\ }\href {\doibase 10.1007/jhep11(2014)161} {\bibfield  {journal} {\bibinfo  {journal} {Journal of High Energy Physics}\ }\textbf {\bibinfo {volume} {2014}} (\bibinfo {year} {2014}),\ 10.1007/jhep11(2014)161}\BibitemShut {NoStop}%
\bibitem [{\citenamefont {Hollik}(2016)}]{Hollik_2016}%
  \BibitemOpen
  \bibfield  {author} {\bibinfo {author} {\bibfnamefont {W.~G.}\ \bibnamefont {Hollik}},\ }\href {\doibase 10.1007/jhep08(2016)126} {\bibfield  {journal} {\bibinfo  {journal} {Journal of High Energy Physics}\ }\textbf {\bibinfo {volume} {2016}} (\bibinfo {year} {2016}),\ 10.1007/jhep08(2016)126}\BibitemShut {NoStop}%
\bibitem [{\citenamefont {Staub}(2019)}]{Staub:2018vux}%
  \BibitemOpen
  \bibfield  {author} {\bibinfo {author} {\bibfnamefont {F.}~\bibnamefont {Staub}},\ }\href {\doibase 10.1016/j.physletb.2018.12.039} {\bibfield  {journal} {\bibinfo  {journal} {Phys. Lett. B}\ }\textbf {\bibinfo {volume} {789}},\ \bibinfo {pages} {203} (\bibinfo {year} {2019})},\ \Eprint {http://arxiv.org/abs/1811.08300} {arXiv:1811.08300 [hep-ph]} \BibitemShut {NoStop}%
\bibitem [{\citenamefont {Duan}\ \emph {et~al.}(2019)\citenamefont {Duan}, \citenamefont {Han}, \citenamefont {Peng}, \citenamefont {Wu},\ and\ \citenamefont {Yang}}]{Duan:2018cgb}%
  \BibitemOpen
  \bibfield  {author} {\bibinfo {author} {\bibfnamefont {G.~H.}\ \bibnamefont {Duan}}, \bibinfo {author} {\bibfnamefont {C.}~\bibnamefont {Han}}, \bibinfo {author} {\bibfnamefont {B.}~\bibnamefont {Peng}}, \bibinfo {author} {\bibfnamefont {L.}~\bibnamefont {Wu}}, \ and\ \bibinfo {author} {\bibfnamefont {J.~M.}\ \bibnamefont {Yang}},\ }\href {\doibase 10.1016/j.physletb.2018.12.001} {\bibfield  {journal} {\bibinfo  {journal} {Phys. Lett. B}\ }\textbf {\bibinfo {volume} {788}},\ \bibinfo {pages} {475} (\bibinfo {year} {2019})},\ \Eprint {http://arxiv.org/abs/1809.10061} {arXiv:1809.10061 [hep-ph]} \BibitemShut {NoStop}%
\bibitem [{\citenamefont {Hollik}\ \emph {et~al.}(2019{\natexlab{a}})\citenamefont {Hollik}, \citenamefont {Weiglein},\ and\ \citenamefont {Wittbrodt}}]{Hollik:2018wrr}%
  \BibitemOpen
  \bibfield  {author} {\bibinfo {author} {\bibfnamefont {W.~G.}\ \bibnamefont {Hollik}}, \bibinfo {author} {\bibfnamefont {G.}~\bibnamefont {Weiglein}}, \ and\ \bibinfo {author} {\bibfnamefont {J.}~\bibnamefont {Wittbrodt}},\ }\href {\doibase 10.1007/JHEP03(2019)109} {\bibfield  {journal} {\bibinfo  {journal} {JHEP}\ }\textbf {\bibinfo {volume} {03}},\ \bibinfo {pages} {109} (\bibinfo {year} {2019}{\natexlab{a}})},\ \Eprint {http://arxiv.org/abs/1812.04644} {arXiv:1812.04644 [hep-ph]} \BibitemShut {NoStop}%
\bibitem [{\citenamefont {Hollik}\ \emph {et~al.}(2019{\natexlab{b}})\citenamefont {Hollik}, \citenamefont {Liebler}, \citenamefont {Moortgat-Pick}, \citenamefont {Paßehr},\ and\ \citenamefont {Weiglein}}]{Hollik_2019}%
  \BibitemOpen
  \bibfield  {author} {\bibinfo {author} {\bibfnamefont {W.~G.}\ \bibnamefont {Hollik}}, \bibinfo {author} {\bibfnamefont {S.}~\bibnamefont {Liebler}}, \bibinfo {author} {\bibfnamefont {G.}~\bibnamefont {Moortgat-Pick}}, \bibinfo {author} {\bibfnamefont {S.}~\bibnamefont {Paßehr}}, \ and\ \bibinfo {author} {\bibfnamefont {G.}~\bibnamefont {Weiglein}},\ }\href {\doibase 10.1140/epjc/s10052-019-6561-6} {\bibfield  {journal} {\bibinfo  {journal} {The European Physical Journal C}\ }\textbf {\bibinfo {volume} {79}} (\bibinfo {year} {2019}{\natexlab{b}}),\ 10.1140/epjc/s10052-019-6561-6}\BibitemShut {NoStop}%
\bibitem [{\citenamefont {Abel}\ and\ \citenamefont {Savoy}(1998)}]{Abel:1998ie}%
  \BibitemOpen
  \bibfield  {author} {\bibinfo {author} {\bibfnamefont {S.~A.}\ \bibnamefont {Abel}}\ and\ \bibinfo {author} {\bibfnamefont {C.~A.}\ \bibnamefont {Savoy}},\ }\href {\doibase 10.1016/S0550-3213(98)00450-7} {\bibfield  {journal} {\bibinfo  {journal} {Nucl. Phys. B}\ }\textbf {\bibinfo {volume} {532}},\ \bibinfo {pages} {3} (\bibinfo {year} {1998})},\ \Eprint {http://arxiv.org/abs/hep-ph/9803218} {arXiv:hep-ph/9803218} \BibitemShut {NoStop}%
\bibitem [{\citenamefont {Staub}(2014)}]{Staub:2013tta}%
  \BibitemOpen
  \bibfield  {author} {\bibinfo {author} {\bibfnamefont {F.}~\bibnamefont {Staub}},\ }\href {\doibase 10.1016/j.cpc.2014.02.018} {\bibfield  {journal} {\bibinfo  {journal} {Comput. Phys. Commun.}\ }\textbf {\bibinfo {volume} {185}},\ \bibinfo {pages} {1773} (\bibinfo {year} {2014})},\ \Eprint {http://arxiv.org/abs/1309.7223} {arXiv:1309.7223 [hep-ph]} \BibitemShut {NoStop}%
\bibitem [{\citenamefont {Porod}\ and\ \citenamefont {Staub}(2012)}]{Porod:2011nf}%
  \BibitemOpen
  \bibfield  {author} {\bibinfo {author} {\bibfnamefont {W.}~\bibnamefont {Porod}}\ and\ \bibinfo {author} {\bibfnamefont {F.}~\bibnamefont {Staub}},\ }\href {\doibase 10.1016/j.cpc.2012.05.021} {\bibfield  {journal} {\bibinfo  {journal} {Comput. Phys. Commun.}\ }\textbf {\bibinfo {volume} {183}},\ \bibinfo {pages} {2458} (\bibinfo {year} {2012})},\ \Eprint {http://arxiv.org/abs/1104.1573} {arXiv:1104.1573 [hep-ph]} \BibitemShut {NoStop}%
\bibitem [{\citenamefont {Coleman}(1977)}]{Coleman:1977py}%
  \BibitemOpen
  \bibfield  {author} {\bibinfo {author} {\bibfnamefont {S.~R.}\ \bibnamefont {Coleman}},\ }\href {\doibase 10.1103/PhysRevD.16.1248} {\bibfield  {journal} {\bibinfo  {journal} {Phys. Rev. D}\ }\textbf {\bibinfo {volume} {15}},\ \bibinfo {pages} {2929} (\bibinfo {year} {1977})},\ \bibinfo {note} {[Erratum: Phys.Rev.D 16, 1248 (1977)]}\BibitemShut {NoStop}%
\bibitem [{\citenamefont {Kane}\ \emph {et~al.}(1994)\citenamefont {Kane}, \citenamefont {Kolda}, \citenamefont {Roszkowski},\ and\ \citenamefont {Wells}}]{Kane:1993td}%
  \BibitemOpen
  \bibfield  {author} {\bibinfo {author} {\bibfnamefont {G.~L.}\ \bibnamefont {Kane}}, \bibinfo {author} {\bibfnamefont {C.~F.}\ \bibnamefont {Kolda}}, \bibinfo {author} {\bibfnamefont {L.}~\bibnamefont {Roszkowski}}, \ and\ \bibinfo {author} {\bibfnamefont {J.~D.}\ \bibnamefont {Wells}},\ }\href {\doibase 10.1103/PhysRevD.49.6173} {\bibfield  {journal} {\bibinfo  {journal} {Phys. Rev. D}\ }\textbf {\bibinfo {volume} {49}},\ \bibinfo {pages} {6173} (\bibinfo {year} {1994})},\ \Eprint {http://arxiv.org/abs/hep-ph/9312272} {arXiv:hep-ph/9312272} \BibitemShut {NoStop}%
\bibitem [{\citenamefont {Bahl}\ \emph {et~al.}(2020)\citenamefont {Bahl}, \citenamefont {Heinemeyer}, \citenamefont {Hollik},\ and\ \citenamefont {Weiglein}}]{Bahl_2020}%
  \BibitemOpen
  \bibfield  {author} {\bibinfo {author} {\bibfnamefont {H.}~\bibnamefont {Bahl}}, \bibinfo {author} {\bibfnamefont {S.}~\bibnamefont {Heinemeyer}}, \bibinfo {author} {\bibfnamefont {W.}~\bibnamefont {Hollik}}, \ and\ \bibinfo {author} {\bibfnamefont {G.}~\bibnamefont {Weiglein}},\ }\href {\doibase 10.1140/epjc/s10052-020-8079-3} {\bibfield  {journal} {\bibinfo  {journal} {The European Physical Journal C}\ }\textbf {\bibinfo {volume} {80}} (\bibinfo {year} {2020}),\ 10.1140/epjc/s10052-020-8079-3}\BibitemShut {NoStop}%
\bibitem [{\citenamefont {Staub}\ and\ \citenamefont {Porod}(2017)}]{Staub_2017}%
  \BibitemOpen
  \bibfield  {author} {\bibinfo {author} {\bibfnamefont {F.}~\bibnamefont {Staub}}\ and\ \bibinfo {author} {\bibfnamefont {W.}~\bibnamefont {Porod}},\ }\href {\doibase 10.1140/epjc/s10052-017-4893-7} {\bibfield  {journal} {\bibinfo  {journal} {The European Physical Journal C}\ }\textbf {\bibinfo {volume} {77}} (\bibinfo {year} {2017}),\ 10.1140/epjc/s10052-017-4893-7}\BibitemShut {NoStop}%
\bibitem [{\citenamefont {Guada}\ and\ \citenamefont {Nemev\v{s}ek}(2020)}]{Guada:2020ihz}%
  \BibitemOpen
  \bibfield  {author} {\bibinfo {author} {\bibfnamefont {V.}~\bibnamefont {Guada}}\ and\ \bibinfo {author} {\bibfnamefont {M.}~\bibnamefont {Nemev\v{s}ek}},\ }\href {\doibase 10.1103/PhysRevD.102.125017} {\bibfield  {journal} {\bibinfo  {journal} {Phys. Rev. D}\ }\textbf {\bibinfo {volume} {102}},\ \bibinfo {pages} {125017} (\bibinfo {year} {2020})},\ \Eprint {http://arxiv.org/abs/2009.01535} {arXiv:2009.01535 [hep-th]} \BibitemShut {NoStop}%
\end{thebibliography}%

\end{document}